# Reversible Ionic Liquid Intercalation for Electrically Controlled Thermal Radiation from Graphene Devices

Xiaoxiao Yu[1,2], Gokhan Bakan[1,2], Hengyi Guo[1], M. Said Ergoktas[1,2], Pietro Steiner[1,2], Coskun Kocabas*[1,2,3]

1  Department of Materials, The University of Manchester, M13 9PL Manchester, United Kingdom
2  National Graphene Institute, The University of Manchester, M13 9PL Manchester, United Kingdom
3  Henry Royce Institute for Advanced Materials, Royce Hub Building, The University of Manchester, M13 9PL Manchester, United Kingdom

Corresponding Author: coskun.kocabas@manchester.ac.uk

**Abstract**

Using graphene as a tuneable optical material enables a series of optical devices such as switchable radar absorbers, variable infrared emissivity surfaces, or visible electrochromic devices. These devices rely on controlling the charge density on graphene with electrostatic gating or intercalation. In this paper, we studied the effect of ionic liquid intercalation on the long-term performance of optoelectronic devices operating within a broad infrared wavelength range. Our spectroscopic and thermal characterization results reveal the key limiting factors for the intercalation process and the performance of the infrared devices, such as the electrolyte ion-size asymmetry and charge distribution scheme and the effects of oxygen. Our results provide insight for the limiting mechanism for graphene applications in infrared thermal management and tunable heat signature control.

TOC FIGURE:

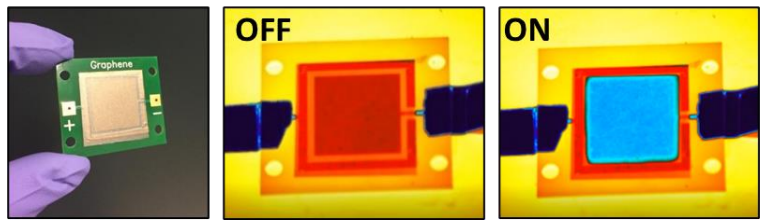

Graphene-based optoelectronic devices can control and manipulate electromagnetic waves over a broad spectrum.[1] Switchable radar absorbers,[2] tunable infrared surfaces[3, 4] or electrochromic devices operating in the visible regime[4-7] are some of these emerging devices that exploit tunable optical properties of graphene. Because of the hexagonal crystal and the linear electronic band structures of graphene[1] as well as the versatility for processing and fabrication, graphene-based materials are broadly investigated as a tunable material for optoelectronics, *e.g.*, photodetectors[8] and optoelectronic modulators.[6] Various approaches are being investigated for changing the charge density and the Fermi energy[7, 9-12] to control the optical properties. Methods like surface charge transfer or substitutional doping leave the material with fixed charge and even alter its intrinsic structure.[7] A few approaches reported to manipulate the optical properties of materials include injection or depletion of substrate carriers, photoexcitation of surface carriers, and electrolyte gating.[13, 14] Among these tuning methods, electrostatic gating and ion intercalation could be electrically tunable and reversible processes that allow for controlling the physical properties of host materials in the infrared wavelength range.[3, 7, 10] Rather than having charged species only residing at the surface electrodes in electrostatic gating, intercalation offers significantly larger doping concentration for thick materials as the ions could migrate into the interlayers of a host material.[15, 16] For these purposes, the host material is desired to be a conductive thin surface that enables the intercalation capacity. And this makes the multilayer graphene (MLG) films a promising choice, owing to their high electrical conductivity and broadband tunability.

Early research on intercalated graphite compounds, mainly motivated by the energy storage applications, investigated the intercalation of metal ions and inorganics into graphitic materials, where the staging phenomena and tunable physical properties have been observed.[3, 4, 7, 10, 15] However, the reactive nature of metal ions, especially lithium ions, and the associated electrolyte medium critically limit the operating conditions and thus require an inert environment and proper sealing in the fabrication process.[4] More recently, other intercalants such as room-temperature ionic liquids (ILs) are also confirmed as possible for device applications.[17-19] The rich variety of ionic liquids and the ability to engineer the functional and peripheral groups can enable optimization of the specific device performance for emerging applications.

In this work, we investigate the broad range of room temperature ionic liquids to optimize the performance of devices operating in the infrared wavelengths. These devices operate as tunable emissivity surfaces. The thermal radiation for graphene films can be controlled by the applied voltage which drives the intercalation of ions. The range of emissivity modulation, Fermi energy shift, and long-term stability of these devices critically depend on the type of ionic liquid. The doping mechanism and electrochemistry of the intercalation process remain elusive. We investigated a few parameters that can potentially affect the performance of such devices, in terms of the selection criteria for ILs, the specification of MLG, and the controlling voltage range. Hence, it is possible to optimize the thermal emissivity modulation of the devices. Voltage-dependent infrared reflection measurements, X-ray diffraction, and thermal imaging characterization during the cycling of the devices reveal the key mechanism behind the gating of the graphene layer and the electrochemical stability of devices. We observed that oxidation of doped graphene under ambient conditions is the main limiting factor for the long-term stability of devices. We show that hole-doped graphene achieved by anion intercalation is significantly more stable than electron-doped (cation-intercalated) graphene. We also found that coating graphene with a thin layer of an oxygen diffusion barrier significantly enhances the long-term stability. We have also observed that the relative size and charge distribution difference between the anion and cation of the ionic liquid determine the voltage drop across the device, playing an important role of the intercalation onset and device operation.

Undoped graphene is a broadband optical absorber mediated by the interband and intraband transition of electrons.[2] Doped graphene, however, has an optical gap in the absorption spectrum due to the Pauli blocking of interband transitions.[2, 15, 20] The location of Fermi energy determines the onset of this gap ($\frac{hc}{\lambda} < 2E_F$), where $h$ is the Planck constant, $c$ is the speed of light in a vacuum, $\lambda$ is the wavelength of the absorption onset, and $E_F$ is the Fermi energy.[4, 20] To understand the tunable optical properties of graphene devices, we first need to analyze the opposing behaviors of interband and intraband transitions as the Fermi energy shifts with the doping. As the electron density increases, the Fermi energy shifts up to 1 eV. Therefore, MLG becomes optically transparent starting from far-infrared and moving into the visible wavelengths at higher doping.[7, 20] On the other hand, intraband transitions provide a Drude-like metallic response, which reflects long-wavelength light. At higher charge densities, graphene becomes more metallic, resulting in higher reflectivity at the infrared wavelengths. These two mechanisms enable the voltage-controlled variable infrared emissivity. Although the absorption of single-layer graphene is insignificant,[20] for practical applications we need to enhance the absorption by stacking the graphene layers which could reach >80% absorption in the infrared wavelengths. The charge density of such MLG films can be tuned by intercalation of ions.[15] And the infrared optical response of electrons in graphene can be described by the Drude model through the frequency-dependent optical conductivity.[20]

Controlling the infrared absorption of graphene enables controlling the thermal radiation on demand. Thermodynamically, at thermal equilibrium, the optical absorption ($\alpha$) and emission ($\varepsilon$) are identical. If an object absorbs at a wavelength, it should radiate at the same time; this is known as Kirchhoff's law of radiation. On a macroscopic perspective, the total energy radiated per unit area from a surface is described by the Stefan−Boltzmann law, such that $P = \varepsilon\sigma T^4$, where $\varepsilon$ is the surface thermal emissivity, $\sigma$ is the Stefan−Boltzmann constant, and $T$ is the actual surface temperature.[3] However, the thermal radiation from the device also includes the reflection of the thermal background, which could be significant in a closed lab environment. Therefore, $P = \varepsilon\sigma T_{act}^4 + R\sigma T_{amb}^4$, where $T_{act}$ is the device physical temperature and $T_{amb}$ is the background temperature of the room. Therefore, by monitoring the apparent temperature of a surface using an infrared camera, it is viable to obtain the emission and absorption, and thus the emissivity and the modulation depth, of the device. Hence, we can evaluate the performance of the ionic liquid electrolyte with one of the consistent standards, the tuning range of thermal emissivity, and have it compared with the ones from the literature.

Our work focuses on understanding the principles within the intercalation process and optimizing the thermal modulation range of such devices and their long-term stability. Furthermore, the devices are investigated regarding the preservation or improvement in operating conditions, lifetime, and the applicable wavelength range. A robust device of this kind should provide fast switching in optical responses, and thus could be applied for realistic applications such as thermal management of satellites or tunable control of heat signatures.

## Results and Discussion

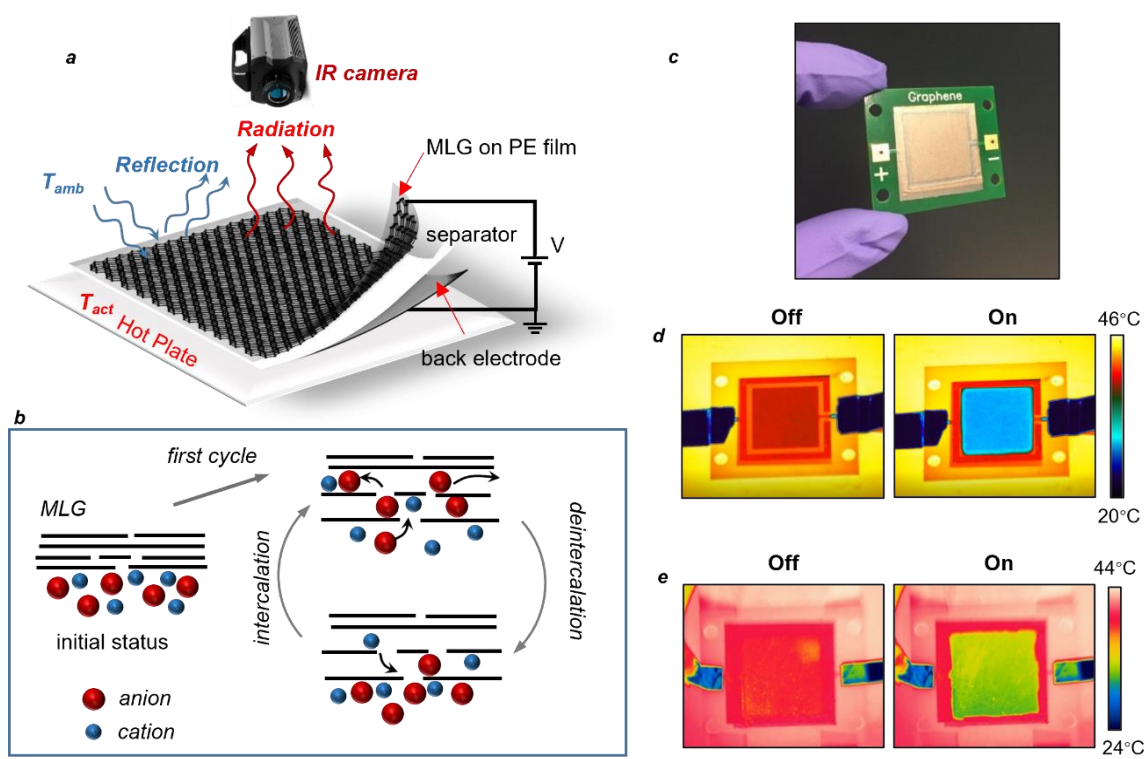

**Figure 1. (a)** Schematic structure of device consisting of laminated layers of MLG film, an electrolyte layer in the polyethylene membrane, and a back electrode. The thermal characterization setup includes an LWIR or MWIR camera, a hot plate, and a voltage source. **(b)** Schematic drawing of the mechanism of ionic liquid intercalation of a multilayer graphene film initiated by the defects at the grain boundaries followed by diffusion of ions between the layers. **(c)** Photograph of the infrared device fabricated on a printed circuit board using gold-plated electrodes laminated by electrolyte and graphene layers. **(d, e)** Thermal camera images of the device at high and low emissivity states recorded by (d) MWIR (3–5 μm) and (e) LWIR (8–14 μm) cameras. Broad tunable absorption of graphene enables modulation of apparent temperature both at mid- and long-infrared wavelength.

**Infrared Characterization of the Devices**. Figure 1 a presents a structure of the graphene infrared device and a comprehensive sketch of the overall monitoring setup using an IR camera. The device is fixed on a hot plate to ensure good heat conductivity. The device consists of four layers: the back electrode (on a printed circuit board), the separating membrane soaked with IL, the MLG film, and the IR transparent top protection layer (Figure 1 a). The MLG layer is prepared by laminating 20-μm-thick low-density polyethylene (LDPE) on top of the MLG. This PE layer is to reinforce the brittle nature of MLG itself and to optimize the handling processes while allowing maximal thermal transmission under the operating wavelength. A bias voltage is applied to the device using a source-measure unit (Keithley 2400) and recorded altogether with the electric current during the intercalation and deintercalation steps. A thermal image of the device is recorded *in situ* using the infrared camera (FLIR T660), from which the average apparent temperature, and thus the average surface emissivity, of the device is extracted to correlate the voltage and emissivity data sets. To extract the correct emissivity of the device in the long-wavelength infrared, we used two reference samples (aluminum foil and polyamide tape) that enable real-time thermal references for the hot plate temperature and background thermal reflection. Using the aforementioned Stefan−Boltzmann law and the heat balance at the device surface, it is feasible to associate the device thermal emissivity $\varepsilon$ with measurable variables as $T_{app}^4 = \varepsilon T_{act}^4 + (1-\varepsilon)T_{amb}^4$, where $T_{app}$, $T_{act}$, and $T_{amb}$ represent the apparent temperature of the device surface, actual temperature of the hot plate, and ambient temperature, correspondingly. All temperatures are recorded by the infrared camera, where the $T_{act}$ and $T_{amb}$ are calibrated by polyimide tape ($\varepsilon = 0.9$) and aluminum foil ($\varepsilon = 0.02$), respectively (Supporting Materials Figure S1).

Schematic profiles of MLG during intercalation cycles are demonstrated in Figure 1 b, where the gaps represent defects and grain boundaries in graphene layers that naturally occur during the CVD growth process. Despite of the natural defects, low sheet resistance was measured for various batches of MLG in the range ~50−200 Ω/sq, and the grain size of CVD-synthesized MLG can vary from 0.1 to 100 μm.[21] Hence the ions would prefer to access through these gaps as their sizes range from ~0.4 to ~1 nm in length,[22, 23] which is much larger than the graphene intraplanar lattice spacing (~0.335 nm).[24] The intercalation process was also monitored under a thermal camera with a close-up lens of 25 μm spatial resolution. It shows that intercalation begins from the defects (Supporting Materials Videos S1 and Figure S8) followed by an in-plane diffusion. We also observed dual intercalation of anions and cations. Under a positive bias voltage, the MLG film becomes hole-doped due to anion insertion; however, *in situ* XPS spectra confirm presence of cations with a charge imbalance.[3] Intercalated ions then expand the interlayer spacing between graphene sheets and lead to an irreversible structural change of MLG.

The device on a standard printed circuit board (PCB) substrate optimized for infrared characterizations is demonstrated as in Figure 1 c. Figure 1 d,e exhibit the representative thermal images of the device at high and low emissivity recorded with mid-wavelength (3−5μm) and long-wavelength (7−14μm) IR cameras. Controlling the bias voltage allows the device to switch between the on and off states under a broad infrared range (Supporting Materials Videos

S2 and Video S3). Corresponding to the analytical derivations, the apparent temperature can be monitored *in situ* to track the thermal emissivity dynamically as a function of the bias voltage.

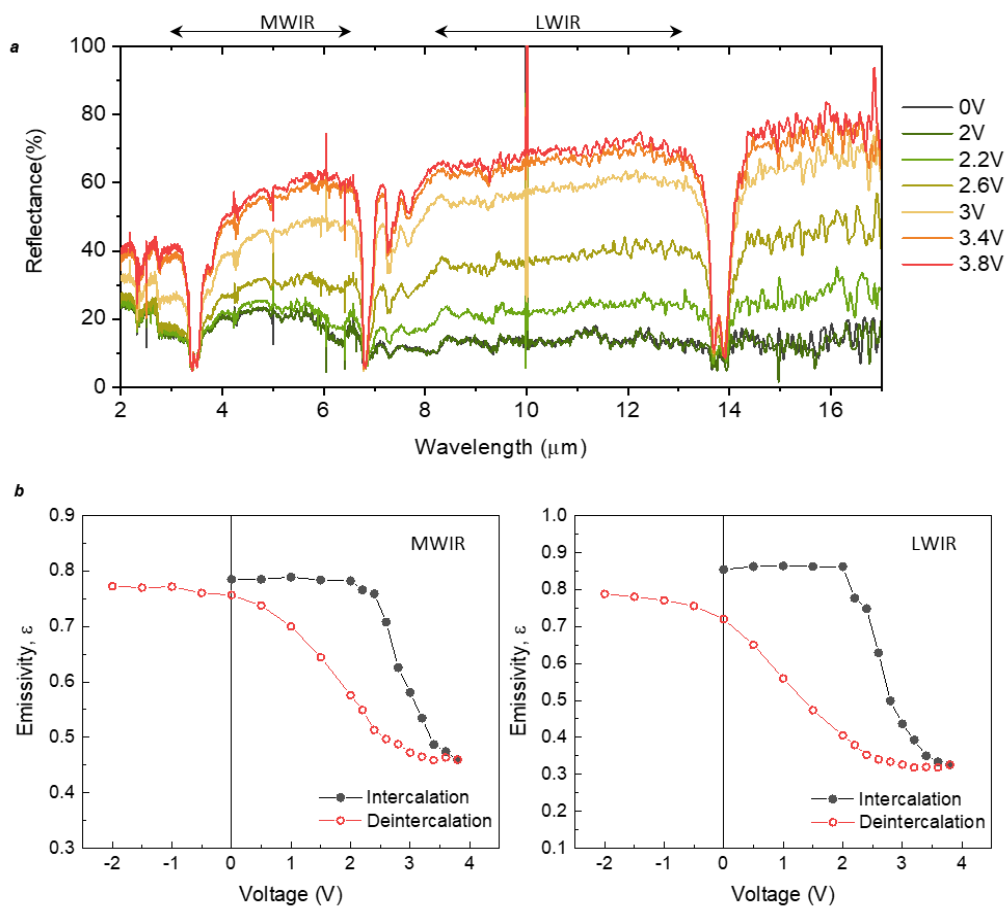

**Figure 2.** Spectroscopic characterization of an MLG infrared device. **(a)** *In situ* Fourier-transform infrared spectroscopy (FTIR) characterization of a device at increasing voltage (from 0 to 3.8 V). **(b)** Voltage dependence of thermal emissivity of the surface for MWIR (3–5 μm) and LWIR (8–12 μm) ranges and with stage recovery after intercalation.

The features of IL intercalation and thus the device performance are further characterized by an infrared spectrometer equipped with an integrating sphere and cooled MCT detector as shown in Figure 2 a. Each spectrum is corrected using a near-perfect absorber and reflector calibration samples. To obtain the best electrochemical stability, we fabricated the devices using a stainless steel (SS) back electrode. The separator, a lens cleaning tissue, is wetted by 1-allyl-3-methylimidazolium bis(trifluoromethylsulfonyl)imide ([AMIM][TFSI], 99%, Iolitec) electrolyte. The measurement is carried out *in situ* to track the maximum reflectance, and thus the minimum emissivity, that the device achieves at each voltage level.

The results verify that the reflectance of the IL-based device is reversibly tunable through a broad range of infrared spectra, from 2 to 17 μm. This is applicable to nearly the entire MWIR and LWIR ranges, except at the fingerprint absorbing wavelengths of the cover layer LDPE (at approximately 3.4, 6.8, 13.7, and 13.9 μm). Fourier-transform infrared spectroscopy (FTIR) characterization also points out that the threshold voltage of intercalation for the particular [AMIM][TFSI]-based device is 2.2 V. The threshold voltage is understood as a critical point in the first intercalation cycle when electrical double layers are formed on both the back electrode and graphene surface. As the CVD method provides MLG with relatively consistent high electric conductivity (~50−200 Ω/sq), defect density and grain sizes are not tuned in this study. We focused on parameters potentially affecting the voltage drop across the device and observed that the threshold voltage depends on the material and specific surface area of the back electrode, the size of ionic liquid particles, and the thickness of the ionic liquid layer. This intercalation process is observed to be reversible for both MWIR and LWIR, as presented in Figure 2 b. To attempt to restore the device to its initial state, the bias voltage decreases step by step following the intercalation process, from 3.8 V to −2 V. The deintercalation process leads to an emissivity recovery of 98.4% in MWIR and 92.3% in LWIR, for the first cycle of charging/discharging (Supporting Information Figure S2 provides details of the deintercalation process). It is interesting to note that the device can maintain the low emissivity state even when the voltage source is disconnected.

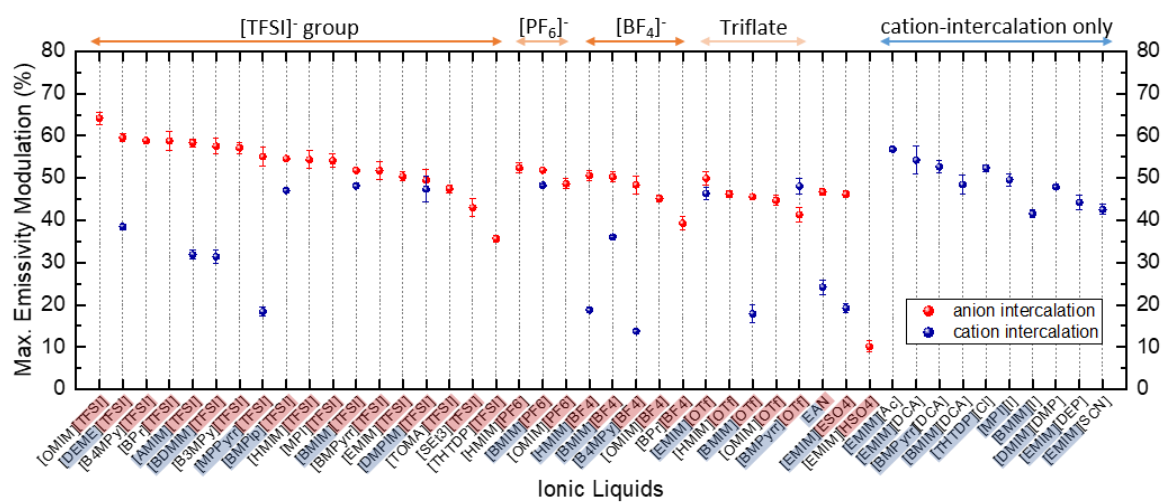

**Figure 3.** Graph showing the measured LWIR emissivity modulation for 44 different ionic liquids providing anion or/and cation intercalation. The ILs are grouped by common anions such as $TFSI^-$, $PF_6^-$, and $BF_4^-$. Red and blue colors represent anion and cation intercalation, respectively. And error bars are plotted with standard deviations of three different measurements of each device.

**Investigation of the Ionic Liquids.** We have investigated 44 different commercial ionic liquids and measured the infrared emissivity modulation using the setup shown in Figure 1 a. All devices were fabricated with identical materials except for the IL electrolyte. MLG used for device fabrication is from the same CVD batch and cut into square sheets of ~2 cm × 2 cm to keep the consistency between devices. Figure 3 provides the measured emissivity modulation grouped by the common anion and cation. The devices were intercalated by increasing the bias voltage up to ±4.0 V. Most of the ILs yield emissivity modulation at positive voltages with anion intercalation. Sixteen ILs are capable of intercalating both the anion and the cation, regardless of the modulation efficiency. [TFSI]$^-$ performs as the highest emissivity modulation intercalant when paired with other cations, in terms of large emissivity modulation and lower voltages. Ten ILs show only cation intercalation when a large cation is paired with a small anion such as Cl$^-$ or I$^-$. This corresponds with the ionic size asymmetry and is related to the ion charge distribution scheme.[25-27] Interestingly, we observe that intercalation of smaller ions usually requires a larger voltage. We were not able to intercalate very small anions such as Cl$^-$ and I$^-$ within the voltage range (±5 V).

**Capacitance Model of the Device**. Experimental and computational methods have been investigated by many groups to understand the electrical double layer (EDL) between the IL and electrode surface.[28-30] *In situ* atomic force microscopy (AFM) on a flat Au plate verifies that the cation in 1-butyl-1-methylpyrrolidinium tris(pentafluoroethyl)trifluorophosphate ([Py$_{1,4}$]FAP) can be tightly restrained to gold surface at only a −2.0 V surface potential. Accordingly, we have developed an electrostatic model, shown in Figure 4 a, to understand the effect of ionic size asymmetry, which directly affects the threshold voltage. The device can be modeled with two EDL capacitors associated with the electrode−IL interface and IL−graphene interface. Assuming that the capacitors are in series, the voltage drop across these capacitors is inversely proportional to the interface EDL capacitance, $\Delta V = \frac{Q}{C}$. Here, the EDL capacitance can be written as $C = \frac{\varepsilon_0 \varepsilon}{d}$ where $\varepsilon$ is the dielectric constant and $d$ is the thickness of the EDL. Since there is no solvent in the IL, the thickness of the EDL is directly related with the size of the ions. Smaller ions yield larger EDL capacitance, leading to a smaller voltage drop at the interface. The force driving the intercalation into graphene layers is the voltage drop across the graphene−electrolyte interface. This voltage drop is proportional to the size of the ion. We have verified the capacitance model by measuring the bulk voltage of the electrolyte for seven different ILs. We observe that depending on the ion size the bulk IL voltage changes drastically (Supporting Materials Figure S3).

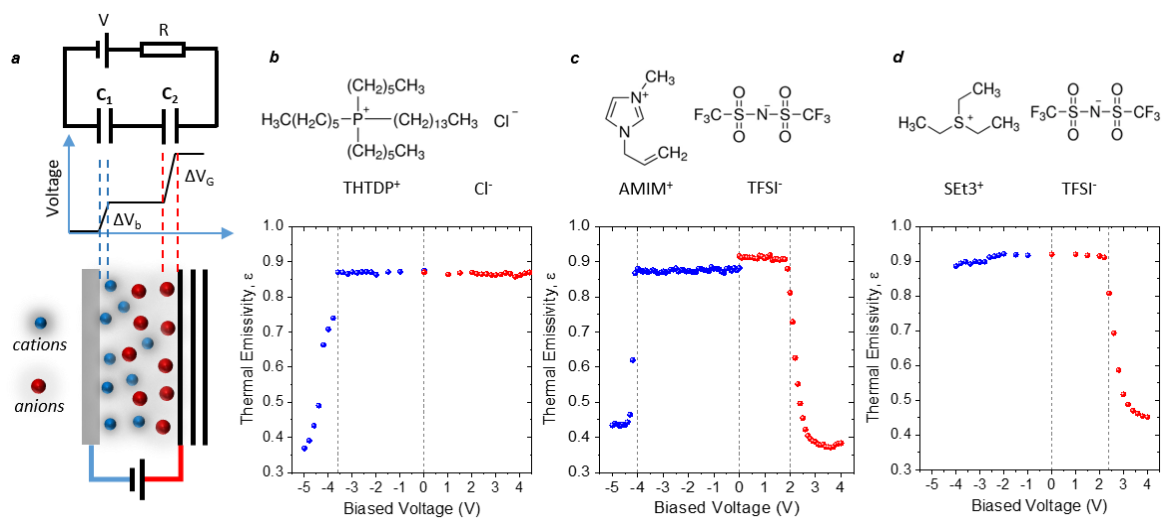

**Figure 4. (a)** The electrostatic model of the device represented by two electrical double layer capacitors for each interface. The voltage drop across the device is inversely proportional to the EDL capacitor for each interface. The voltage of the bulk ionic liquid is determined by the balance between these capacitances. **(b–d)** The three cases of ion size asymmetry and different charge distribution schemes. (b) Large cation (THTDP$^+$) and small anion (Cl$^-$) both with a concentrated charge distribution. (c) Cation (AMIM$^+$) and anion (TFSI$^-$) have a concentrated and delocalized charge, respectively, but with equivalent ion size. (d) Small cation (SEt3$^+$, concentrated charge) and large anion (TFSI$^-$, delocalized charge distribution). The graphs show the modulation of the emissivity as a function of voltage for each case.

**Table 1.** Threshold Voltages and Key Emissivity Results from [AMIM][TFSI] Intercalation and Deintercalation Measurement.

| intercalated ion | threshold voltage, $V_{th}$ | initial emissivity, $\varepsilon_0$ | minimum emissivity, $\varepsilon_{min}$ | restored emissivity, $\varepsilon_r$ | recovery |
|---|---|---|---|---|---|
| [TFSI]$^-$ | 2.0 V | 0.92 | 0.37 | 0.84 | 91.3% |
| [AMIM]$^+$ | −4.2 V | 0.88 | 0.43 | 0.83 | 94.3% |

To support this model, we show the voltage dependent emissivity for three cases: (1) large cation (THTDP$^+$) paired with small anion (Cl$^-$), both with concentrated charge distribution; (2) anion and cation with comparable ionic size; cation (AMIM$^+$) with concentrated charge and anion (TFSI$^-$) with delocalized charge distribution; and (3) small cation (Set3$^+$) paired with relatively large anion (TFSI$^-$) with similar charge distribution scheme to [AMIM][TFSI]. For the first case, we observed only the intercalation of a large cation at −3.8 V (Figure 4 b). For the second case (Figure 4 c), we observed intercalation of both anion and cations at −4.2 and 2 V, respectively. The third case (Figure 4 d) shows intercalation of only the large anion [TFSI]$^-$. From these observations, we can draw the conclusion that the intercalation threshold depends on the ion size. Counterintuitively, the threshold voltage required to intercalate a large ion is relatively small when it is paired with a smaller counterion. When the size of the ions is comparable, the threshold voltage is likely determined by the interaction of the ions with the surface (*i.e.*, interface capacitance).

Therefore, further investigation regarding the deintercalation features of both cation and anion is presented in Table 1. Two devices fabricated in the same way were charged at opposite biased voltages to intercalate and deintercalate the anions and cations individually. Both devices were tested for the first cycle of voltage scanning. And the minimal thermal emissivity achieved at each voltage level was noted as in Figure 4 c. Results are summarized in Table 1. Both the cation and the anion exhibit good thermal emissivity modulation (55% for [TFSI]$^-$ and 45% for [AMIM]$^+$) in the first scanning cycle. However, the exceptionally high threshold voltage of [AMIM]$^+$ intercalation is disadvantageous, as it may lead to side redox reactions with the electrolyte or at the MLG surface. Moreover, a large voltage drop at the electrode surface requires a higher threshold voltage to initiate the intercalation process. This could lead to a short device lifetime in the cyclic endurance test. And in spite of generally large electrochemical windows of ILs, a high $V_{th}$ is not desired for durable applications due to the active nature of current collecting materials and energy consumption.

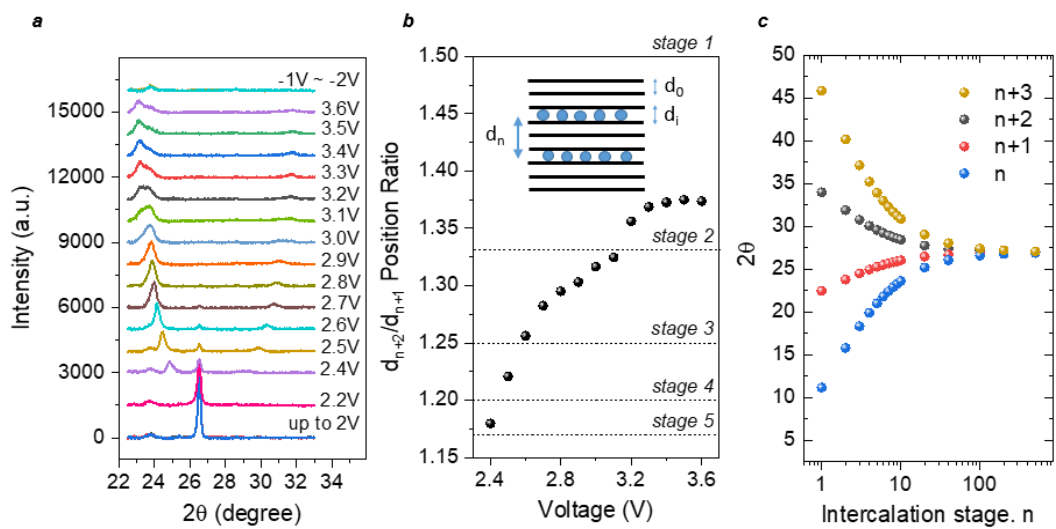

**Figure 5.** *In situ* XRD characterization and analysis of the [AMIM][TFSI]-based device. **(a)** XRD results of charging the device (intercalated with [TFSI]⁻) from 0 to 3.6 V, followed by discharging at up to −2 V. **(b)** Peak analysis refers to the stage of intercalation achieved at a varied voltage level. The inset shows the schematic of the stage 3 intercalated graphene layers. **(c)** Theoretical higher-order peak positions of [TFSI]⁻ intercalation at varied intercalation stages.

*In Situ* **Recording of X-ray Diffraction from the Devices.** To further understand the intercalation process, we discuss the effect of intercalation on the structural changes in MLG inspected with *in situ* X-ray diffraction (XRD). Figure 5 a summarizes the results of *in situ* XRD measurements obtained from a device during the charging cycle. The initial dominant peak at 2θ =26.5° represents the (002) plane of the graphitic structure.[31] As the voltage increases, the original graphene stacking structure is gradually disrupted by intercalating the [TFSI]⁻ anion, so that the intensity of the (002) diffraction diminishes. New diffraction peaks associated with the periodic intercalated planes emerge as the interplanar spacing between the intercalated layers as $d_n = d_i + (n-1)d_0$ ; here, $d_0$ is the interatomic distance between the planes of graphite and $d_i$ is the thickness of the intercalated layer.[31] These correspond with results in the literature and refer to the primary (00 n+1) and secondary dominant peak (00 n+2) where n is the intercalation stage.[31, 32] The ratio between the two peak positions determines the intercalation stage of a graphite intercalation compound, as demonstrated in Figure 5 b. The theoretical (00 n+1) and (00 n+2) peak positions of [TFSI]⁻ intercalation are derived from intercalation stages and plotted in Figure 5 c. The values at nearly stage 2 agree with the experimental XRD results in Figure 5 a. XRD results denote that the (002) peak completely vanishes at above 2.8 V. Unlike the emissivity, the structural deformation caused by the intercalation process is irreversible likely due to the residual ions between the layers. It is reported that stage 1 intercalation of [TFSI]$^-$ is approachable at elevated bias voltage and temperature.[31] The results suggest that mild voltages and a higher intercalation stage could bring less damage to the MLG and extend the lifetime of devices.

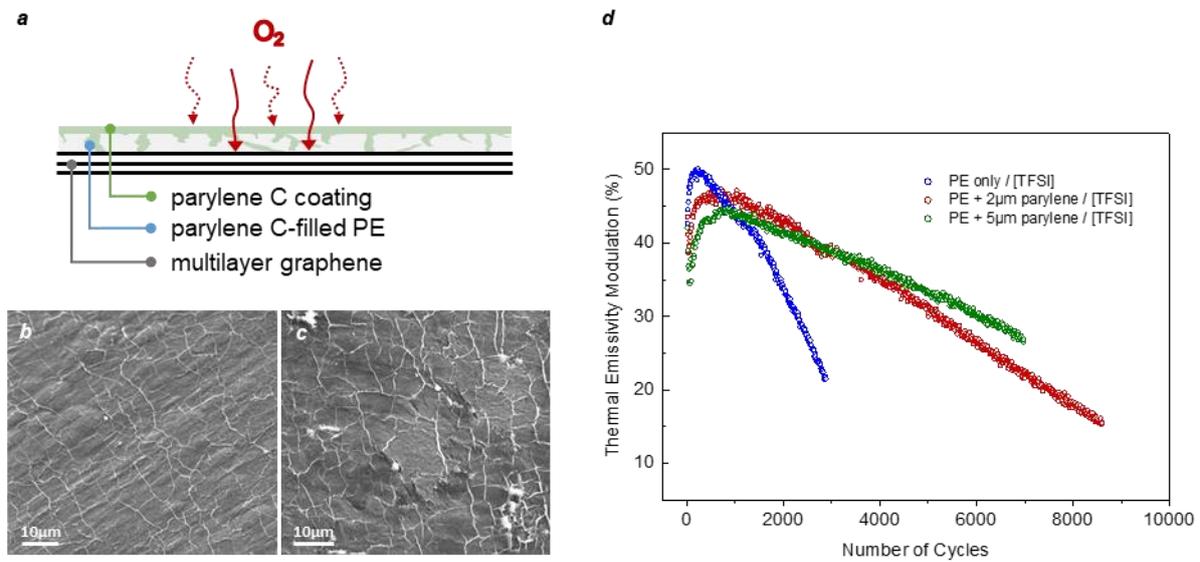

**Figure 6.** Further modification and characterization of the top MLG layer, and endurance tests on modified devices. **(a)** Schematic of Parylene C coating on top of the LDPE to reduce oxygen permeability in the device. **(b, c)** Scanning electron microscopy (SEM) images of a pristine MLG surface (b) and an MLG surface after cyclic tests (c). **(d)** Effects of varied Parylene C coating thickness on the device thermal modulation depth. Devices consist of SS, tissue soaked with [AMIM][TFSI], and MLG layer.

**Long Term Stability of the Devices**: For realistic applications, the long-term reversible cycling of the devices plays a critical role. To distinguish factors that limit the lifetime of devices, we recorded long-term thermal imaging of the devices under cyclic voltages between 2.8 and −1.5 V. We observed that the emissivity modulation initially shows an improvement. However, after a few hundred cycles, the performance of the device decays significantly. We noticed that this degradation is because of the oxidation of the graphene layer by the oxygen molecules diffusing through the PE overcoating. To prevent oxygen diffusion, the top layer was first coated with Parylene C by the CVD method to restrain oxygen from contacting the MLG. As presented in Figure 6 a, a thin layer of conformal Parylene C deposition coats the porous LDPE film[33] to enhance the oxygen barrier of MLG while maintaining high infrared transmittance (Supporting Materials Figure S4). Devices prepared for cyclic endurance tests were fabricated with MLG from the same CVD batch to eliminate the impacts of device-to-device variation. Figure 6d shows the results of cyclic endurance tests of three devices synthesized with no, 2 μm, and 5 μm Parylene C coating. The devices were switched on and off by holding 3s at 2.8 V and −1.5 V, correspondingly and repetitively. Cyclic tests show that devices coated with Parylene C from 5 μm to none attain the maximal thermal modulation depth of 45.0%, 47.8%, and 50.2%. And their thermal tuning ability is significantly reduced to less than 30% after *ca.* 6000, 5000, and 2300 cycles, respectively. All devices present a nearly steady degeneration rate after achieving the maximum modulation. This long-term degradation is likely due to the oxidation of the graphene layer and intrinsic mechanical effects during the intercalation process as well as the redox reaction at the MLG surface (Supporting Materials Figure S5).

Figure 6 b,c exhibit SEM images of the graphene film before and after the intercalation cycles, showing most of the invasive structural change that was introduced to MLG by IL intercalation occurs at defects and grain boundaries. The images show that the intercalation is initiated through the grain and defect boundaries. After thousands of cyclic intercalations and deintercalations, some grains are mechanically detached from the continuous film, leading to electrically isolated islands. The isolated islands cannot be intercalated, resulting in less emissivity modulation of the device.

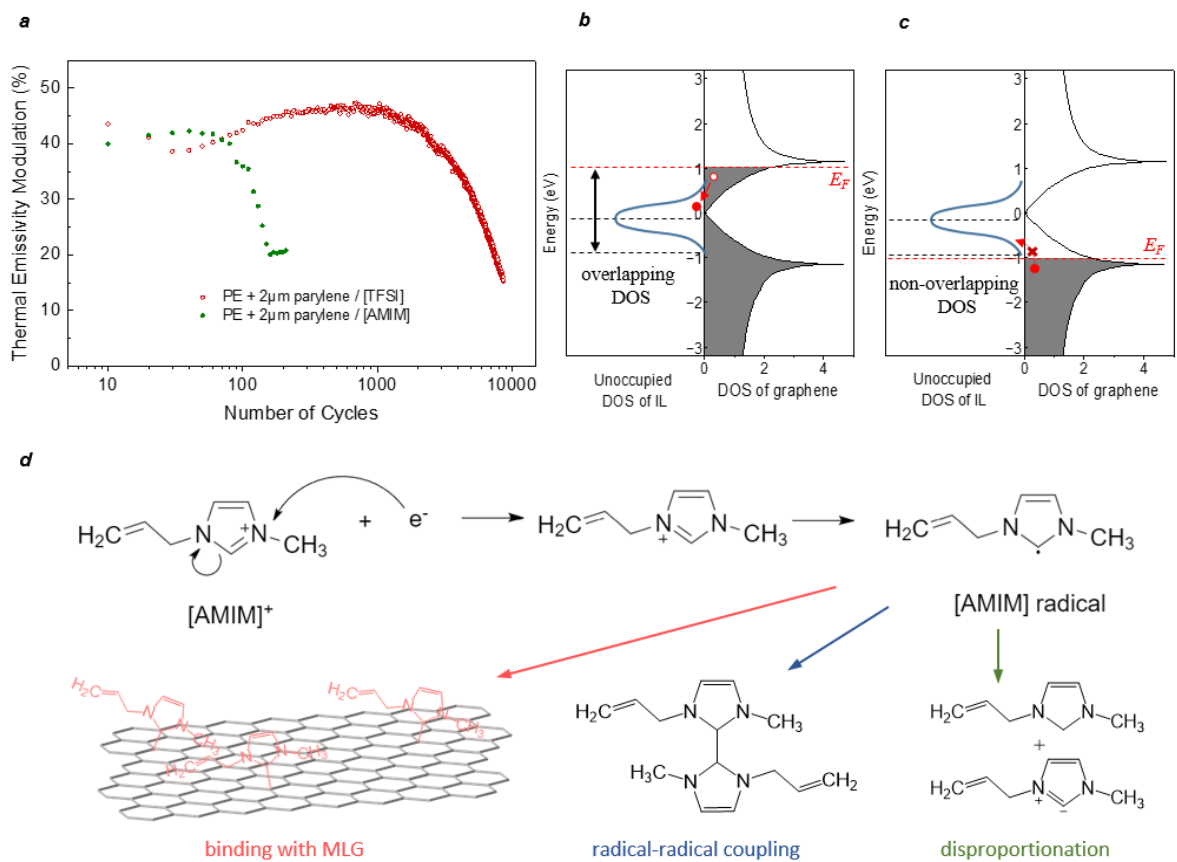

**Figure 7.** Effects of different doping scheme on device oxidation. **(a)** Long term endurance test of the infrared devices showing the stability difference between electron- and hole- doping achieved by intercalation of AMIM and TFSI ions. **(b,c)** Schematic illustration of graphene oxidation principle with density of state (DOS). (b) Overlapping DOS caused by n-type doping is desired for electron transfer from graphene to electrolyte, which oxidises graphene. (c) Lowered Fermi level with p-type doping prevents oxidation on graphene and improves device lifetime. **(d)** Potential pathways of cation deposition on graphene interlayers with side reactions.

Although additional coating of oxygen diffusion barrier enhances the long-term stability of the thermal emissivity modulation, the slope of curves evidently verifies that the MLG with relatively thick and dense coating demonstrates better durability and thus longer lifetime. The endurance tests of a device with alternative IL present similar results. (Supporting materials Figure S6) This also indicates that ambient oxygen content participates in the oxidation of doped MLG.

**Stability of electron or hole doped graphene:** To investigate the effect of the dopant type (hole or electron doping) on the oxidation rate of the graphene, we carried out cyclic test for electron-doped graphene using cation [AMIM]$^+$ intercalation, and for hole-doped graphene using [TFSI]⁻ intercalation. The results presented in Figure 7a show noteworthy difference in device lifetime. The thermal modulation of cation intercalation decreases to 30% after only 100 cycles and falls to 20% after 150 cycles where device is generally inapplicable. These results show that electron-doped graphene is significantly more reactive than hole-doped graphene because the high electron density and high Fermi energy can initiate the electron transfer between graphene and electrolyte. As demonstrated in Figure 7(b,c), the electron-doped MLG leads to elevated Fermi level, thus it tends to lose electrons to IL with lower unoccupied electronic states, leaving graphene prone to oxidation.[21, 34] Hole-doped graphene, on the other hand, reduces its Fermi level, making it difficult for electron transfer. Another reason that cations are not favoured for intercalation is shown in Figure 7d. It also presents a potential reaction pathway during the intercalation of cations. Given by quantum chemical semi-empirical calculations, the cation [AMIM]⁺ could be reduced to [AMIM] radical as it encounters electrons near the MLG layers.[27, 35] The unpaired electron of this radical may bind with a localised electron of the π-bond to deposit [AMIM] radicals on graphene interlayers.[35, 36] Other prevailing pathways where two [AMIM] radicals can either couple with each other to form a dimer or have a disproportionation reaction to decompose the IL.[27] Therefore, charging under relatively high bias voltage for [AMIM]⁺ intercalation is easier to cause electrochemical decomposition of electrolyte. And the MLG is more likely to be oxidised with oxygen, resulting in poor optical switching performance and short device lifetime.

**Conclusion**

As a conclusion, we studied the performance of graphene-based infrared devices with controllable infrared emissivity. We have investigated 44 different ionic liquids and determined the best performing electrolytes. Our results shows that the intercalation of [TFSI]⁻ anions provides the best performance in terms of emissivity modulation and long-term stability. We determined that the oxygen diffusion through the overlayer is the main limiting factor of the endurance of the devices. Using a thin oxygen diffusion barrier significantly improves the long-term stability. Furthermore, we observed that hole-doped devices are significantly more stable than electron-doped ones, because electron-doped graphene is very reactive to oxygen and electrolytes, resulting very short device lifetime. We have also observed that size difference between anion and cations of the ionic liquid determine the threshold voltage for the intercalation. We provide a capacitance model explained by the asymmetric voltage drop at the interface between the electrode-electrolyte. Our results provide feasible approaches with mechanism understanding for graphene-based infrared emissivity devices with potential applications for thermal management and adaptive camouflage applications.

**Materials and Methods**

*Chemical vapor deposition of MLG.* MLG was synthesized by chemical vapor deposition on Ni foils of 25μm thickness (Alfa Aesar, 12722). The Ni foil was heated to 900°C under 100 sccm $H_2$ flow and 100 sccm Ar flow, and annealed at 900°C for 20 min. Afterwards, it was treated under 50 sccm $CH_4$ flow for 15 min at atmospheric pressure, followed by 100 sccm $H_2$ flow and 100 sccm Ar flow at 900°C. Finally, the sample was cooled down to room temperature under 100 sccm $H_2$ flow and 100 sccm Ar flow.

*Device fabrication.* The 50 μm-thick stainless steel back electrode was used as received (Agar Scientific, 43-200). It was connected with stainless steel wire for voltage supply. The back electrode was covered by a piece of porous separator used as received from Cytiva (Whatman 105 lens cleaning tissue, 2105-841), which was then soaked with ionic liquid received from Iolitec. The top layer was fabricated by laminating 20μm-thick polyethylene on the MLG-Ni foil at 160°C. The MLG was then easily peeled off from Ni foil and transferred to PE. The PE-MLG top layer was gently placed and flattened on the wetted separator with stainless steel wire connected at MLG side. The functional area of three layers overlapped and adhered together by the surface tension of ionic liquid. Additional Parylene C coating was achieved by CVD on top of the PE-MLG layer prior to device assembly. The sheet resistance of peeled-off MLG on PE was obtained by the four-point resistance measurement using Keithley 2110 multimeter.

*Parylene C coating.* The PE-MLG layer was used as the substrate for all Parylene C coating procedures. The MLG side and sample edge were sealed and protected by polyamide tape on a flat plastic board. The desired thickness of Parylene C coating was determined by weight of initial dimer (dichloro-p-cyclophane, DPX-C, Galentis Ltd) and achieved by a SCS PDS 2010 Labcoter system. The furnace was first heat up to 690°C. The deposition chamber was vacuumed before vaporizer being switched on. After the vaporizer reaching 175°C, both the vaporizer and furnace were switched off to cool down.

*Thermal Camera characterisation.* Thermal images and videos were captured by a thermal camera (FLIR, T660) with setup shown in Figure 1a. The ambient temperature and actual temperature were measured from the surface of Al foil and polyamide tape, respectively. Both were captured in video recordings. The close-up video (Video S1) was recorded with a close-up IR lens (FLIR, T198000). Voltage supply for the device was provided by a Keithley 2400 Sourcemeter. The timelines of thermal camera and sourcemeter were synchronized for real-time recording.

***Fourier-transform infrared spectroscopy.*** FTIR measurement of infrared reflectance was carried out with Perkin Elmer Spectrum 100 FTIR spectrometer equipped with an integrating sphere (PIKE Mid-IR IntegratIR) and a wide-band liquid-nitrogen-cooled mercury-cadmium-telluride detector at a spectral resolution of 2 cm$^{-1}$. Power supply to the device was provided by a Keithley 2400 Sourcemeter.

***X-ray Diffraction characterisation.*** The XRD characterisation was carried out *in situ* with a Rigaku Smartlab X-ray diffractometer, which generates X-rays with Cu K α source. Step bias voltage was applied to the device at the beginning of each XRD scan to the sample device through a Keithley 2400 sourcemeter. The applied voltage increased from 0V until 3.6V during the intercalation process, followed by deintercalation at -1V and -2V. The evolution of peaks at corresponding bias voltage are presented in Figure 5.

***Scanning electron microscopy.*** The SEM characterisation results in Figure 6(b-c) were carried out with Tescan Mira3 SC. The PE-MLG layer after cyclic test was gently cleaned with isopropanol and deionized water to remove the residue ionic liquid and vacuum-dried at 60°C prior to characterisation. The *in-situ* SEM characterisation in Figure S7 was achieved by FEI Quanta 200. The electrically conductive port allowed it to manipulate bias voltage applied to the device while taking SEM images. Power supply was provided by a Keithley 2400 Sourcemeter.

**Supporting Information**

Real-time recordings of the intercalation-deintercalation process using LWIR (Video S1 and Video S3) and MWIR (Video S2) cameras. Infrared characterisations for the calibration samples including the aluminium foil, polyamide tape, and Parylene-C layer (Figure S1 and Figure S4). Additional data for the infrared characterisation of the devices (Figure S2). The capacitive model and the static voltage drop measurement for different ionic liquid samples (Figure S3). XPS analysis of the oxidised graphene after long-term cycling (Figure S5). The intercalation mechanism was investigated through *in situ* SEM (Figure S7) and close-up thermal camera measurement (Figure S8). Details related to FTIR characterisation regarding the deintercalation process and cycling performance of device with other IL are available in the Supporting Information.

**Acknowledgments**

This research is supported by the European Research Council through an ERC Consolidator Grant (grant no. 682723, SmartGraphene) and an ERC PoC Grant (grant no. 899908, SmartIR).